\documentclass[%
reprint,
superscriptaddress,
amsmath,amssymb,
aps,
prb,
floatfix,
]{revtex4-2}

\usepackage{graphicx}
\usepackage{dcolumn}
\usepackage{bm}
\usepackage{braket}
\usepackage{soul}
\usepackage[colorlinks]{hyperref}
\usepackage{color}
\usepackage[usenames,dvipsnames]{xcolor}
\usepackage{comment}
\usepackage{amsmath,amssymb}

\emergencystretch=\maxdimen
\newcommand{\mw}[1]{#1}

\newcommand{\be}{\begin{equation}}
\newcommand{\ee}{\end{equation}}

\begin{document}

\small

\title{Low-noise parametric microwave amplifier based on self-heated nonlinear impedance with sub-nanosecond thermal response}

\author{Marco Will}
\affiliation{Low Temperature Laboratory, Department of Applied Physics, Aalto University, P.O. Box 15100, FI-00076 Espoo, Finland}
\affiliation{QTF Centre of Excellence, Department of Applied Physics, Aalto University, P.O. Box 15100, FI-00076 Aalto, Finland}

\author{Mohammad Tasnimul Haque}
\affiliation{Low Temperature Laboratory, Department of Applied Physics, Aalto University, P.O. Box 15100, FI-00076 Espoo, Finland}
\affiliation{QTF Centre of Excellence, Department of Applied Physics, Aalto University, P.O. Box 15100, FI-00076 Aalto, Finland}

\author{Yuvraj Chaudhry}
\affiliation{Low Temperature Laboratory, Department of Applied Physics, Aalto University, P.O. Box 15100, FI-00076 Espoo, Finland}
\affiliation{QTF Centre of Excellence, Department of Applied Physics, Aalto University, P.O. Box 15100, FI-00076 Aalto, Finland}

\author{Dmitry Golubev}
\altaffiliation[Present address: ]
{HQS Quantum Simulations GmbH, Karlsruhe, Germany}
\affiliation{QTF Centre of Excellence, Department of Applied Physics, Aalto University, P.O. Box 15100, FI-00076 Aalto, Finland}

\author{Pertti Hakonen}
\affiliation{Low Temperature Laboratory, Department of Applied Physics, Aalto University, P.O. Box 15100, FI-00076 Espoo, Finland}
\affiliation{QTF Centre of Excellence, Department of Applied Physics, Aalto University, P.O. Box 15100, FI-00076 Aalto, Finland}

\date{\today}

\begin{abstract}
 Low-noise amplifiers are of great importance in the field of quantum technologies. We study a thermally driven parametric amplifier based on a \mw{superconductor-insulator-graphene-insulator-superconductor (SIGIS)} junction coupled to a superconducting microwave cavity. The strong non-linearity in the temperature dependence of our device leads to \mw{thermal self-modulation that produces impedance oscillations at frequencies around twice the angular cavity resonance frequency $\omega_\mathrm{r}$}. \mw{In particular,} reactance modulation of the effective capacitance yields \mw{a gain of 18.6 dB over a frequency span of 125 kHz with a minimum noise temperature of $T_\mathrm{N} = 1.4$ K}. Our theoretical modelling gives insight into the exact mixing processes, confirmation of the electron-phonon coupling parameter and possible improvements of the studied system.
\end{abstract}

\maketitle


\section{Introduction}

Monolayer graphene has extraordinary thermal properties \cite{Pop2012} and it can provide extremely fast thermal response to electrical and optical signals \cite{Fay2011, Graham2012,Fong2012,Betz2012PRL,Aamir2021}. Fast thermal response of graphene can also be employed for making non-linear electronic devices for producing efficient THz detectors by mixing \cite{Karasik2011}. Recently, such mixers operating at low \mw{temperature} $T$ were built using graphene on SiC \cite{LaraAvila2019}, in which thermal relaxation at low bath temperature $T_0$ takes place by diffusion of charge carriers since the electron-phonon coupling is weak \cite{Bistritzer2009,Kubakaddi2009,Clerk2010,Viljas2010,Song2012b,Betz2012NP,Fong2012,Laitinen2014,Virtanen2014,ElFatimy2019,Kokkoniemi2020,Tomi2021}. Regular hot electron behavior can be employed for explaining the underlying electrical transport phenomena. An operation band width of 20\,GHz is projected for an optimized device near the quantum limit. Besides passive mixing, proper thermal modulation of non-linear graphene junction characteristics can be employed for parametric gain.

Low-noise parametric amplifiers are necessary for high-quality microwave measurements of quantum circuits \cite{Clerk2010}. Earlier, self-pumped  SQUID-based parametric amplifiers were used (pumped by Josephson oscillation governed by DC bias), but their noise performance was limited by the dissipation generated by the bias  \cite{Kuzmin1979}. Later, microwave rf-SQUIDs \cite{Ryhanen1989}, evolved into modern, nearly quantum limited superconducting parametric amplifiers \cite{Yurke1988}, in which the non-linear inductance is provided either by Josephson element (JPA) \cite{Yamamoto2008,Bergeal2010,Elo2019} or kinetic inductance of a superconducting line (KPA) \cite{Zmuidzinas2012}. With the development of traveling wave parametric amplifiers (TWPA), even entanglement generation over large bands has been achieved \cite{Perelshtein2022,Esposito2022,Qiu2023}. Still, the technology development for parametric amplifiers is not at all complete, and new principles are regularly searched for \mw{\cite{Massel2011,Lahteenmaki2012,Jebari2018,Deshmukh2022,Renard2022,Higginbotham2023,Xu2023,Kraglund2024}.}

The bolometric mixer of Ref. \citenum{LaraAvila2019} was based on weak localization $\log(T)$-type of temperature dependence. Superconductor-Graphene-Superconductor junctions with poor interfacial transparency (SIGIS, where I refers to "insulating" interfaces)  and negligible Josephson current can give much stronger exponential $T$-dependence, and thus better detection sensitivity can be expected. In this work, we investigate the possibility of using suspended Molybdenum-Rhenium (MoRe)-graphene-MoRe junctions. Our main emphasis is on generation of rapid thermal modulation \mw{within} a parametric microwave cavity \mw{setting} and on understanding the parametric gain produced by the thermal pump signal. In addition, we investigate the influence of thermal self-oscillation and injection locking by signal frequency on the performance of our parametric microwave amplifier.

The basic non-linear element in our work is a superconductor-insulator-normal metal (SIN/NIS) junction that is embedded into a microwave cavity. The SIN element has non-linear behavior both as a function of voltage and temperature (see Supplementary Information (SI), \mw{Sect. IX}). Our device utilizes non-linearity in the temperature dependence, which is invoked by Joule heating tuned to proper average temperature, at which the cavity frequency becomes sufficiently modulated at double frequency of the pump. Our device is partly self-pumped as it generates itself the $2\omega/2\pi \sim 10$\,GHz frequency which is used in the three wave mixing operation of the amplifier. The efficient "thermal pumping" of the cavity frequency at 10\,GHz is facilitated by the extremely fast thermal response of graphene around $1.5-2$\,K at which our SIGIS modulation element works.   When the modulation range of the cavity frequency exceeds its line width, gain is produced. 

In our amplifier scheme, there is no need to have a mechanical resonator element nor a flux-pumped Josephson device for achieving gain, just a strongly temperature dependent \mw{impedance $Z_{\mathrm{SIN}}(T)$ obtained using a normal conductor directly connected to a superconducting microwave cavity. Temperature modulation leads to sufficient capacitive reactance modulation of the normal-superconductor interface. } The relevant  temperature modulation range in our work is $1.5-2$\,K, which together with external dissipation staying around $T=20$\,mK, limits the noise temperature of our amplifier to $T_\mathrm{N} \simeq 1.4$\,K. Even lower noise can be achieved by better selection and tuning of the contact material. 

We \mw{can achieve an exponentially temperature-dependent resistance $\mathcal{R}(T)$} by using a graphene membrane (G) weakly tunnel-coupled (effectively via an insulating layer (I) \mw{)} to a MoRe superconductor (S) having a large energy gap $\Delta/k_B \simeq 15$\,K. The graphene acts therefore as a SINIS resistance with a strong temperature dependence proportional to \mw{$\mathcal{R}(T) \propto \exp(\Delta/k_B T)$}. Recently, in Ref. \onlinecite{Haque2022}, it was shown that \mw{thermally-induced self-oscillations at frequency $f_{\mathrm{T}}=1 \dots 70$ MHz} can appear in such a setting due to competition between temperature dependent Joule heating \mw{$V_2^2/\mathcal{R}(T)$} by the cavity voltage $V_2$ and the counterbalancing action by electron-phonon coupling (see Fig. \ref{fig:Intro}a).  Similar thermal self-oscillations were observed by  Segev et al. \cite{Segev2007} in a ring resonator with a nonlinear element. In the self-oscillation, the $Q$ factor of the \mw{cavity resonator (see Fig. \ref{fig:Intro}b, for more info on the cavity see \mw{Sect. I} of the SI)} basically jumps between a high value ($Q \sim 5000$) and low value ($Q \sim 20$), leading to relaxation oscillation in the photon number with nearly sawtooth-like time dependence \cite{Haque2022}. This behavior facilitates our amplifier operation at the sideband frequencies \mw{$f_\mathrm{sb}=f_\mathrm{p}+f_\mathrm{T}$}, which we demonstrate experimentally in our work \mw{and $f_\mathrm{p}$ the  cavity pump frequency}. We achieve a noise temperature of $T_\mathrm{N} \simeq 1.4$\,K and \mw{for different operation parameter} a gain of nearly 22\,dB for microwave signals around 5.4\,GHz. By injection locking, a single frequency amplifier with slightly improved \mw{$T_\mathrm{N} \simeq 1.1$\,K  was also realized (see Sect. II of the SI)}.
\section{Modeling}
\begin{figure}[]
\centering
\includegraphics[width=0.99\linewidth]{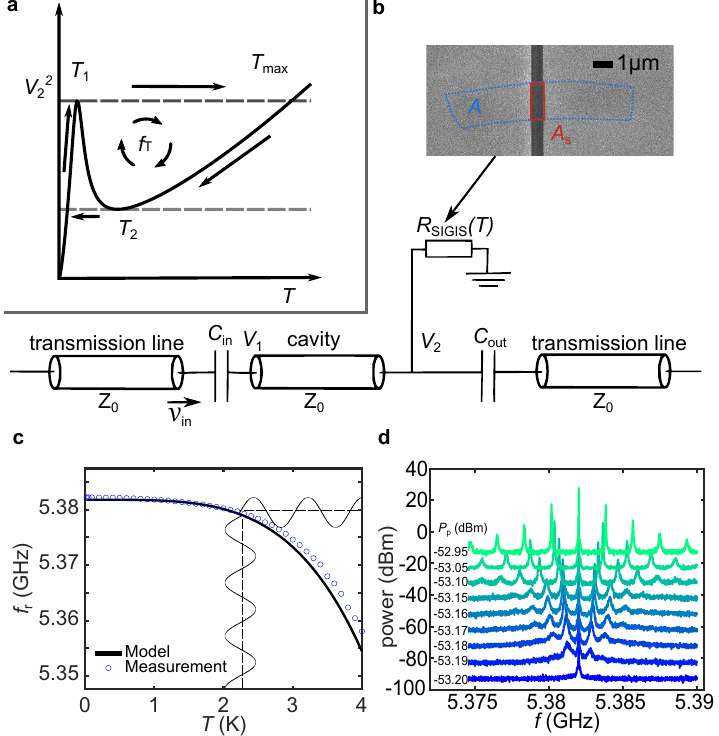}
\caption{General working principle : a) Thermal hysteresis loop of the amplifier with thermal frequency $f_{\mathrm{T}}$ as a function of Joule heating (\mw{voltage drop over graphene} $V_2^2$). b) Scheme of the \mw{coplanar waveguide cavity with graphene SIGIS junction}. The device is pumped with pump power $P_\mathrm{p}$ and frequency $f_\mathrm{p}$ and probed with power $P_{\mathrm{pr}}$ and frequency \mw{$ f_\mathrm{pr}$ via the transmission line and coupling \mw{capacitors $C_\mathrm{in}$ and $C_\mathrm{out}$, respectively}. The cavity voltage $V_1$ and the voltage drop over the graphene $V_2$ are shown in their respective positions.} The inset displays the placement of the total graphene area $A$ and the suspended part \mw{$A_\mathrm{S}$} with respect to the leads. \mw{One electrode is the center conductor of the coplanar waveguide cavity, whereas the other is the ground plane.} c) Cavity resonance frequency $f_\mathrm{r}$ vs. temperature measured without pumping \cite{Haque2022}. With pumping, the cavity frequency decreases due to increased reactance of the SINIS structure as $\delta f_\mathrm{r} \propto 1/\Delta$. The embedded waves illustrate the conversion of temperature wave to frequency modulation.  d) Emission spectra at pump power $P_\mathrm{p} > P_{\mathrm{th}}=-53.2$ dBm. The curves at different $P_\mathrm{p}$ are offset by 10\,dB\mw{m} from each other for visibility.}
\label{fig:Intro}
\end{figure}
According to cavity dissipation experiments at small power \cite{Haque2022}, we may model \mw{the dissipative part of} our SIGIS junction as non-linear resistor $R(T)=1/G(T)$ using an exponential temperature dependence due to variation in quasi-particle density
\begin{equation}
    G(T)= {G_0}+{G_\mathrm{N}} I_0 \left({-\frac{\Delta}{k_BT}} \right), \label{eq:GT}
\end{equation}
where $\Delta\approx1.76k_BT_c$ is the gap of the superconductor, $k_B$ is the Boltzmann constant, $T_c$ the critical temperature of the superconductor, \mw{$T$ can denote the equilibrium temperature or the electronic $T_e$ of the graphene part in non-equilibrium (see \mw{Sect. V} of the SI)}, $G_\mathrm{N}$ is the normal state conductance, $I_0$ is the 0$^{\mathrm{th}}$ modified Bessel function of first kind, and $G_0$ is the effective low temperature conductance due to dissipation of the cavity environment. Here we have assumed for simplicity that the normal conductance is basically limited by one of the contact resistances, i.e., in Eq. \ref{eq:GT} we set $G_N \sim G_\mathrm{NIG}$ with $G_\mathrm{NIG} = 1/R_\mathrm{N}$ \mw{and $R_\mathrm{N}$ the normal state resistance of the graphene}. \mw{Also $R_0=1/G_0$ is the subgap resistance of the SIGIS structure, i.e. its zero bias resistance in the limit $T\to 0$.} Although, we consider here only non-linear $T$ dependence, the conductance depends also non-linearly on the applied voltage (see \mw{Sect. IX of} the SI).

The suspended graphene is coupled to a transmission \mw{coplanar waveguide} microwave cavity with resonance frequency $\mw{f_\mathrm{r}=}\omega_\mathrm{r}(T)/2\pi$. 
Due to reactance of both SIN and NIS interfaces the cavity resonance frequency varies with temperature as shown in Ref. \onlinecite{Haque2022} and depicted in Fig. \ref{fig:Intro}c.
The coupling rate of cavity photons to the graphene ($\kappa_G(T)$), the input port ($\kappa_{\mathrm{in}}$), and the output port ($\kappa_{\mathrm{out}}$) are described by 
\begin{gather}
    \kappa_{G}(T)=\frac{2 \omega_\mathrm{r} Z_0}{\pi R(T)}, \quad  \kappa_{\mathrm{in}}=\frac{2 \omega_\mathrm{r}^3 Z_0^2 C_{\mathrm{in}}^2}{\pi}, \quad \kappa_{\mathrm{out}}=\frac{2 \omega_\mathrm{r}^3 Z_0^2 C_{\mathrm{out}}^2}{\pi}, 
\end{gather}
where $C_{\mathrm{in}}=0.5$\,fF and $C_{\mathrm{out}}=2.2\,$fF are the input and output coupling capacitances, respectively, and $Z_0=100\,\Omega$ is the characteristic line impedance of the resonator. \mw{The total coupling rate is then $\kappa=\kappa_\mathrm{in}+\kappa_\mathrm{out}+\kappa_\mathrm{G}(T_e)$}. The voltage drop over the SIGIS junction $\mw{V_2}$ induced by the pump current $I_{\mathrm{p}}$ and the heat balance of the electron system form a coupled system of equations :
\begin{align}
{2\kappa_{\rm in}}\frac{dv_{\rm in}(t)}{dt}&=
\frac{d^2\mw{V_2}}{dt^2} + \kappa\frac{d\mw{V_2}}{dt} 
+ \frac{2Z_0\omega_\mathrm{r}}{\pi} \frac{d}{dt} 
\left(\frac{\mw{V_2}}{R(T)} \right)
+ \omega_\mathrm{r}^2(t) \mw{V_2},
\label{eq:CSE1} \\ 
C_G(T_e)\frac{dT_e}{dt} &= -\Sigma (T_e^\gamma-T_0^\gamma) + \left( \frac{1}{R(T_e)} + \frac{T_e}{R^2(T_e)} \frac{dR(T_e)}{dT_e} \right)\frac{\mw{V_2}^2}{2},
\label{eq:CSE2}
\end{align}
where the time derivative of \mw{the input voltage is given by a sum of pump and probe signals, $v_{\mathrm{in}}(t) = v_{\mathrm{p}} \sin(\omega_{\mathrm{p}}t)+ v_{\mathrm{pr}} \sin(\omega_{\mathrm{pr}}t)$}, 
\mw{$V_2$ is the voltage drop over the graphene}, and where we have neglected non-linear voltage terms (\mw{see Sect. VI of} the SI); the cavity frequency is time dependent $\omega_\mathrm{r}(T_e (t))/2\pi$ because Joule heating modulates the electronic temperature $T_e$ by doubling the frequency of ac-voltage and even generating mixing products of \mw{$\omega_\mathrm{p}$ and $\omega_{\mathrm{pr}}$}.  In addition, $T_0$ equals the phonon bath temperature, $C_G(T)$ denotes the heat capacity of graphene \cite{Aamir2021}, $A$ is the effective area of the graphene sample, $\Sigma$ denotes the strength of the electron-phonon coupling (EPC) and $\gamma=3$ is the EPC exponent typically observed in the experiments off the Dirac point \cite{Tomi2021}. The heat capacity is given by $C_G(T) = \frac{2Ak_B}{\pi\hbar^2v_F^2}\left( \frac{\pi^2}{3} |\mu| k_B T + \frac{9\zeta(3)}{2}k_B^2T^2 \right) $, where $v_F=10^6$ m/s is the Fermi velocity in graphene and $\mu$ is the chemical potential tunable by gate voltage.
This equation has been experimentally tested \cite{Aamir2021} and found to be accurate.

The Joule heating in graphene is initially compensated by heat transfer by electron-phonon coupling, but due to the exponential increase of the conductance with $T_e$, Joule heating increases faster than what can be compensated by the electron-phonon heat transport. This can lead to a thermal runaway (at $T_1$ in Fig. \ref{fig:Intro}a),  until the heat flow is limited by sample conductance $\sim 1/R_\mathrm{N}$ when the temperature reaches $\simeq \Delta/k_B$ \cite{Haque2022}. Above $T_1$, temperature increases rapidly driving the cavity into a low quality factor regime with large dissipation. This increased dissipation enables quick absorption of photons from the cavity, which leads to reduction of voltage over the junction until the large $Q$-factor regime is reached again (cooling to $T_2$ in Fig. \ref{fig:Intro}a). Fig. \ref{fig:Intro}a shows a schematic of the thermal hysteresis loop. The thermal loop has a frequency $f_\mathrm{T}=\omega_{\mathrm{T}}/2\pi$, which is dependent on the pump power as depicted in Fig. \ref{fig:Intro}d. The threshold for the minimum required pump power below which no thermal loop is formed was determined experimentally at $P_{\mathrm{th}}=-53.2\,$dBm.

In our experiments, the cavity is driven by voltage $v_{\mathrm{in}}(t) = v_\mathrm{p} \sin(\omega_\mathrm{p} t) + v_{\mathrm{pr}} \sin(\omega_{\mathrm{pr}} t)$. \mw{The pump frequency $\omega_\mathrm{p}/2\pi=f_\mathrm{p}$ is selected so that this frequency equals to the cavity resonance at $T=20$\,mK}, while the probe frequency $\omega_\mathrm{pr}/2\pi=f_\mathrm{pr}$ is set to coincide with $f_\mathrm{pr}=f_\mathrm{p}+f_{\mathrm{T}}$.
Because of the rapid thermal response of graphene electrons, the electronic temperature in graphene  can vary at $2f_\mathrm{p}$ under strong pumping. As seen from Eqs. (\ref{eq:CSE1} and \ref{eq:CSE2}), such a \mw{temperature variation $\delta T \propto \sin(2\omega_\mathrm{p} t)$} leads to two extra pumping terms: \mw{two dissipative ones (due to $\kappa_G(T_e)$ and $R(T_e)$) and one reactive (due to $\omega_\mathrm{r}(T_e)$), which are all related to the electronic temperature $T_e$}. The probe signal at the sideband frequency becomes amplified by power transfer from the pump tone. \mw{As there are a lot of different frequencies involved, we list all of them with their relations in Table \ref{tbl:frequencies} in Appendix A.}

\begin{figure}[]
\centering
\includegraphics[width=0.99\linewidth]{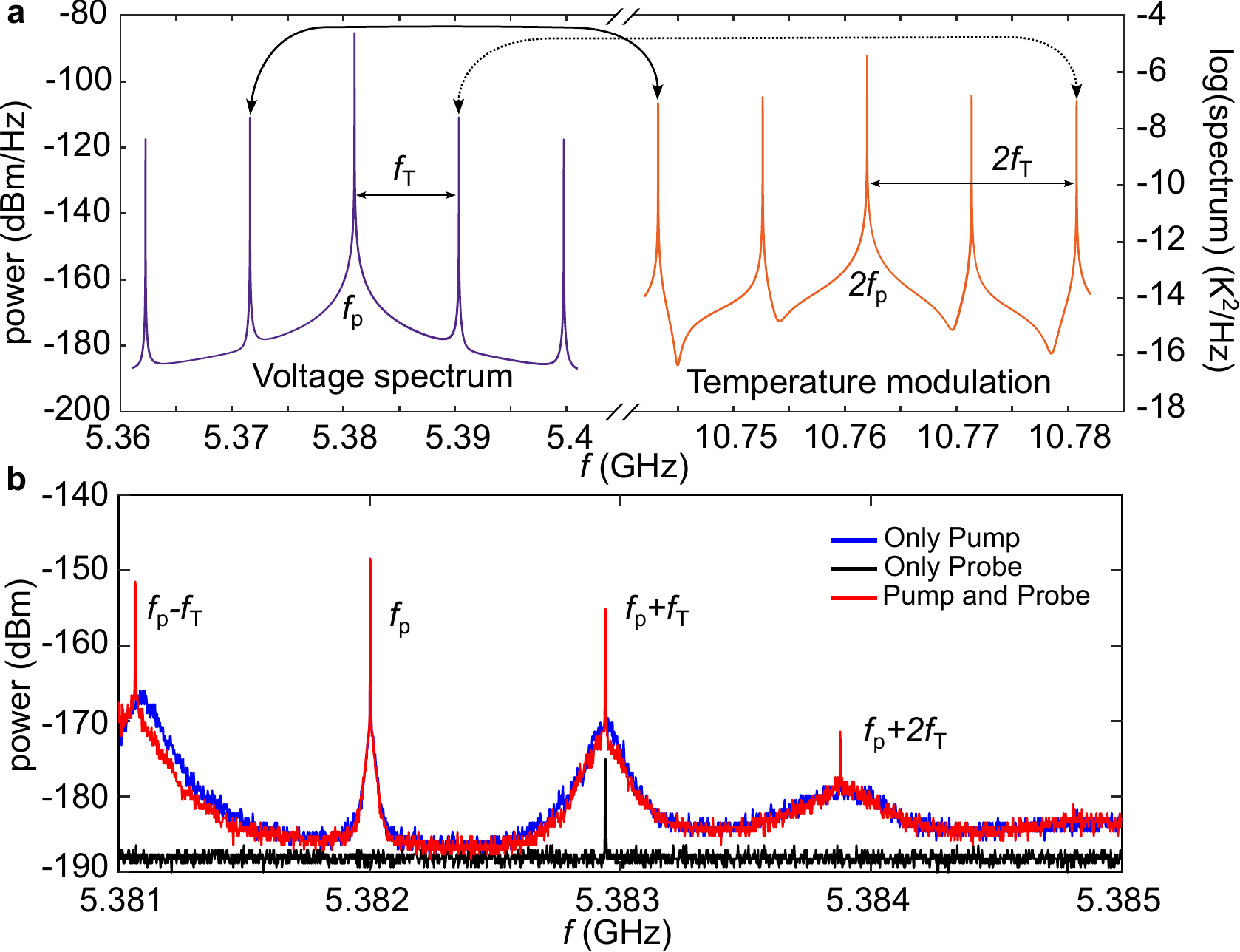}
\caption{\mw{a) Illustration of voltage and temperature spectra calculated using Eqs. \ref{eq:CSE1} and \ref{eq:CSE2}. All frequencies, sideband and cavity, mix together between the voltage and temperature modulated branches. The most important parametric down conversion (PDC) processes in the analytical calculation for the side band frequencies $f_\mathrm{p}+f_{\mathrm{T}}$ and $f_\mathrm{p}-f_{\mathrm{T}}$ are indicated by dashed and solid lines, respectively  (see Sect. VII of the SI)}. 
The displayed curves are from FFT of the simulations with parameters: $A=1.8\,\mu$m$^2$, $\Sigma=2.2\,$WK$^{-3}$m$^{-2}$, $\mu=30$\,meV, $T_c=9\,$K, $R_0=3\cdot 10^5\,\Omega$, $R_\mathrm{N}=500\,\Omega$, $P_\mathrm{p}=-47$\,dBm, \mw{a length of the time trace of} $t_{\mathrm{length}}=100\,\mu$s and 2$^{23}$ sample points. \mw{The variation of these parameters within the self-oscillation regime does not change the characteristic behaviour of the spectra.} The left spectrum is for \mw{voltage} $V_2(t)$ fluctuations, while the right \mw{spectrum is for the $T_e(t)$ fluctuation.}
b) Experimentally measured spectrum around the pump frequency displaying frequencies $f_\mathrm{p}$, $f_\mathrm{p} \pm f_{\mathrm{T}}$, and $f_\mathrm{p} + 2f_{\mathrm{T}}$; \mw{the noise background refers to noise power per Hz}. Note the amplified relative magnitude of the measured sidepeaks at $f_\mathrm{p} \pm f_{\mathrm{T}}$ compared with the pump amplitude.
} 

\label{fig:Mixing}
\end{figure}

In our device, gain may arise either from power flow across frequencies enabled by resistance modulation or by pumping of the cavity frequency. In a parametric system, the relevant power to be amplified is effectively distributed between signal and idler frequencies, which are symmetrically distributed across the pump frequency in our device. Consequently, we look at the power transfer via both signal and idler frequencies. \mw{The sidebands of the pump frequency at $f_\mathrm{p}+f_{\mathrm{T}}$ and at $f_\mathrm{p}-f_{\mathrm{T}}$  cause} a thermal modulation at double frequency $2f_\mathrm{p}+2f_{\mathrm{T}}$ ($2f_\mathrm{p} - 2f_{\mathrm{T}}$), which contributes on top of the temperature oscillation comb caused by the pump and the thermal oscillations. The frequencies $2f_\mathrm{p}+2f_{\mathrm{T}}$ and $2f_\mathrm{p} - 2f_{\mathrm{T}}$ modulate voltage oscillations at $f_\mathrm{p} - f_{\mathrm{T}} $ and $f_\mathrm{p} + f_{\mathrm{T}} $, respectively, which leads to voltage oscillations at $f_\mathrm{p}+f_{\mathrm{T}}$ and $f_\mathrm{p}-f_{\mathrm{T}}$, respectively \mw{(see Fig. \ref{fig:Mixing}a)}. 

\begin{figure}
\begin{centering}
    \includegraphics[width=0.95\linewidth]{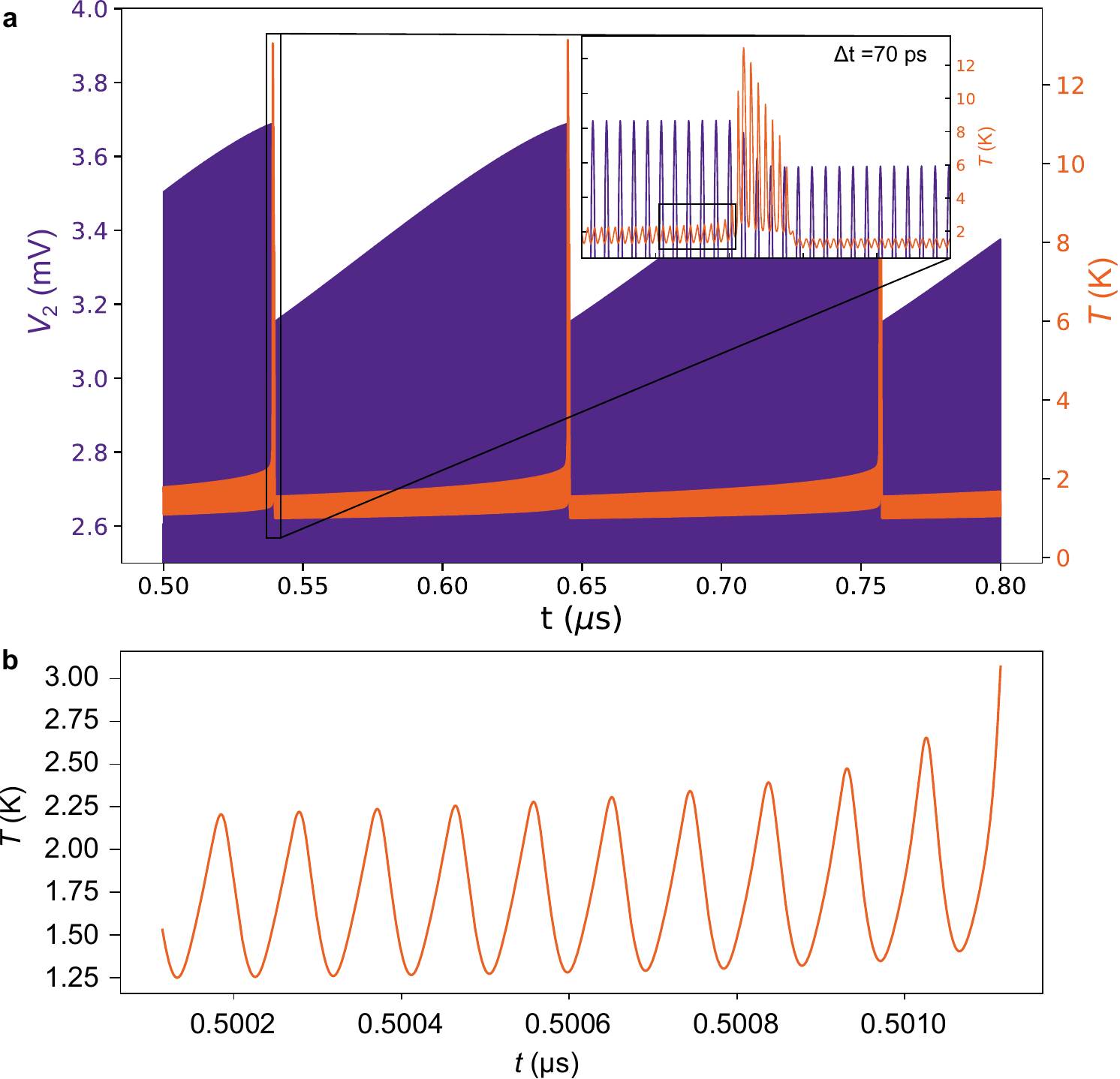}
\caption{
a) Simulated time dependence of temperature $T_e$ and voltage $V_2$ in the  graphene $R_{\mathrm{SIGIS}}$ resistance in the self-oscillating regime. The microwave power is $P_\mathrm{p}=-55$\,dBm and the substrate temperature is $T_0=0.02$\,K; the other parameters are given in the caption of Fig. \ref{fig:Mixing}. The inset shows an expanded view of $T(t)$ and $V_2(t)$ during the 1-ns decay of the cavity photon occupation. \mw{The total time length of the inset is $\Delta t =70$\,ns. }
b) Zoomed-in trace of simulated temperature \mw{of box in inset of} frame a) near the onset of thermal instability. The thermal oscillation around $T_{\mathrm{ave}} \simeq 1.75$\,K amounts approximately to $\delta T_{\mathrm{p-p}} \sim 1$\,K, which \mw{would correspond under equilibrium conditions} to a frequency modulation of the cavity by nearly $2$\,MHz. 
}
\label{fig:TimeDependence}
\end{centering}
\end{figure}

Eqs. \ref{eq:CSE1} and \ref{eq:CSE2} can be solved numerically to obtain the $T(t)$ and $\mw{V_2}(t)$ dependencies \mw{on $v_{\mathrm{in}}(t)$}. Figure \ref{fig:TimeDependence} illustrates the results, which demonstrate the important role of thermal oscillations at $2\omega_\mathrm{p}$. 
Fig. \ref{fig:TimeDependence}a displays the simulated time dependence in the operating regime of our sideband amplifier. The simulation parameters (indicated in the caption of Fig. \ref{fig:TimeDependence}a) set the device into thermal self-oscillation at frequency $f_\mathrm{T}=9$\,MHz. Each thermal run-away is followed by a rapid decrease in the number of photons in the cavity in the high dissipation state (see \mw{Sect. VI of the} SI for analytical modeling with values for damping rates in the high and low $Q$ states). The high dissipation rate decreases \mw{the photon number inside the cavity} $N_\mathrm{r}$ rapidly, and subsequently the cavity starts over to enhance its number of photons in the high $Q$ state. 
The zoomed-in view of $2f_\mathrm{p}$ thermal oscillations around \mw{the time-averaged temperature during the oscillations} $T_{\mathrm{ave}} \simeq 1.7$\,K in Fig. \ref{fig:TimeDependence}b displays a swing of approximately  $\delta T_{\mathrm{p-p}} \sim 1$\,K, 
\mw{which corresponds to a frequency modulation of the cavity by $\sim 2$\,MHz according to the equilibrium dependence of resonance frequency $f_{\mathrm{r}}$ in Fig. \ref{fig:Intro}c}. Note that our numerical simulation neglects details of a SIN junction capacitance \mw{$C_\mathrm{eff}(T,V_2,\nu)$}, the delicate dependencies of which  \mw{are expected to boost the \mw{obtained modulation} ($\nu$ is scaled frequency, see Sect. VII of the SI)}. In fact, the most likely scenario in our experiment is that the voltage is unevenly distributed over two SIN junctions, and that one junction provides the heating while the second junction acts as the modulated reactance.

We have carefully analyzed the dissipation-driven parametric amplification scenario using Fast Fourier Transform (FFT) of the time traces (as illustrated in Figs. \ref{fig:TimeDependence}a and \ref{fig:Mixing}a), but no actual gain could be observed. Thus, we assign the experimentally observed gain to reactive pumping of the cavity frequency at $2\omega_\mathrm{p}$. \mw{The self-oscillation modulated pumping at $2\omega_\mathrm{p}$ leads to parametric down conversion of quanta at $2\omega_\mathrm{p}+2\omega_{\mathrm{T}}$ and $2\omega_\mathrm{p}-2\omega_{\mathrm{T}}$ to sidebands of the pump frequency $\omega_\mathrm{p} \pm \omega_{\mathrm{T}}$, the signal and idler frequencies} (see Fig. \ref{fig:Mixing}b). This cavity frequency modulation can generate amplification, when its swing is more than the line width of the cavity. 
Indeed, the thermal modulation $T(2\omega_\mathrm{p})$ amounting to $T_{e} \simeq 1.7 \pm0.5$\,K according to our simulations indicates that the \mw{frequency modulation} $\delta f(T_e (2\omega_\mathrm{p}))$ can exceed the line width. Moreover, the $2\omega_\mathrm{p}$ modulation of $T_e$ is substantial all along the build-up and decay of oscillations in the cavity: as seen in Fig. \ref{fig:TimeDependence}a, the $2\omega_\mathrm{p}$ temperature swing is even larger during the voltage amplitude collapse, but the duration of this phase of pumping is very short.

At large pump powers, both signal and idler frequencies become amplified owing to PDC. At first sight, the role of thermal self-oscillation is thus to provide a mixer in which power is transferred between different frequencies with good efficiency. However, this mixing has somewhat peculiar characteristics since \mw{ both the real and imaginary parts of the cavity impedance, $\Re( Z(\omega))$ and $\Im (Z(\omega))$,}  as seen by its environment become $\omega$ dependent, and it displays peaks at $\omega_\mathrm{p} \pm k \omega_{\mathrm{T}} $ \mw{as well as at $2\omega_\mathrm{p} \pm k \omega_{\mathrm{T}} $,} where $k$ is an integer. As a consequence, voltage due to PDC processes become boosted beyond the cavity response curve and the range of parametric gain can extend to the side peaks. A regular $2\omega$-pumped parametric cavity may well produce gain over 40\,dB, and if about 20\,dB is lost in mixing (see Fig. \ref{fig:Mixing}), we can still expect 20\, dB of gain from the operated device. 

Typically, low $T$ parametric amplifiers include flux-pumped SQUID loops or SNAILs (superconducting non-linear asymmetric inductive element), using which inductance (cavity frequency) of the circuit is modulated. Our junction does not display any supercurrent, and the non-linear element is provided by high-frequency properties of SIN junctions (see \mw{Sect. IX} of the SI). At voltage \mw{$\alpha V_2 \lesssim 0.5 \Delta/e$, where $\alpha \sim 0.2$ denotes the fraction of $V_2$ across a SIN junction}, the high-frequency response of the SIN junction is capacitive, denoted by $C_{\mathrm{eff}}$, the value of which grows with increasing temperature. Since $C_{\mathrm{eff}}$ is connected in parallel to the equivalent resonant circuit, the frequency decreases with increasing $T$. \mw{The corresponding frequency shift $\delta f_\mathrm{eff}$ can be estimated using $\delta f_{\mathrm{eff}}/f_\mathrm{r} = \frac{1}{2} (\delta C_{\mathrm{eff}}/C) $ where the equivalent cavity capacitance $C=0.3$\,pF (see Sect. \ref{Sect:exp}) and $ \delta C_{\mathrm{eff}}$ is the change in $C_{\mathrm{eff}}$; this yields a relative shift of $ 10^{-3}$ for $ \delta C_{\mathrm{eff}} =0.6$\,fF in our microwave cavity.  To the lowest order (see Sect. IX of the SI), we may write $\delta(\frac{1} {\omega C_{\mathrm{eff}}(T)}) \sim 0.04 (T_{\mathrm{max}}-T_{\mathrm{min}}) R$, with $T_{\mathrm{max}}$ and $T_{\mathrm{min}}$ denoting the maximum and minimum temperature values (in Kelvin) during the modulation, respectively, and setting normal state junction resistance as \mw{$ R=0.2/G_N$}.   
For a low-temperature modulation of 1\,K amplitude, we obtain $\delta C_{\mathrm{eff}} \sim 1.5$\,fF which can produce a substantial frequency shift in the cavity, and thereby generate parametric gain.}
We note, however, that part of the parametric pumping action may arise from the non-linear voltage dependence of the SIN capacitance.

\section{Experiment} \label{Sect:exp}

Fig. \ref{fig:Intro}b displays a schematic view of the measurement and the sample. The \mw{employed $\lambda/2$ microwave cavity in transmission measurement geometry is modelled by a} parallel lumped $LC$ circuit with input ($C_{\mathrm{in}} = 0.5$\,fF) and output capacitors ($C_{\mathrm{out}} = 2.21$\,fF). The equivalent capacitance $C=0.30$ pF and inductance $L= 3.0$ nH yield for the characteristic impedance of the cavity $\sqrt{L/C}=100$ $\Omega$. The graphene piece of size $16 \times 2.5$ $\mu \mathrm{m}^2$ was mounted directly across the 700\,nm gap between the center conductor and the ground plane of a MoRe cavity\mw{. The cavity was} made by etching a sputtered MoRe film of 300\,nm thickness. A lithographically patterned gate electrode was placed under the suspended graphene part for tuning of carrier concentration. 
Details of fabrication and the measurement setting can be found in Ref. \onlinecite{Haque2022} and \mw{in Sect. I} of the SI. 

The coupling between graphene and MoRe was weak, and only negligible proximity effect was achieved, most likely due to small contact area and partial oxidation of the metal surface. No supercurrent was recorded in the studied sample. Exponential increase of conductance with temperature was observed at temperatures $T \gtrsim 2$\,K \cite{Haque2022}, which allows us to identify our graphene as a SINIS device with two superconductor-insulator-normal junctions in series.
Our sample is denoted as resistance $R_{\mathrm{SIGIS}}$ parallel to the cavity circuit. 

In our amplification experiments, we employed pump powers slightly above the threshold power $P_{\mathrm{th}}=-53.2$\,dBm at frequency corresponding to the base temperature resonance \mw{$\omega_0=2\pi\cdot f_0=2\pi\cdot5.382$\,GHz}. 
Pumping at \mw{$\omega_\mathrm{p}=\omega_0$} with powers below the threshold did not lead to the appearance of sidebands. We used a pump power of $P_\mathrm{p}=-53.15$\,dBm for most of our measurements, which yields a separation of $\omega_{T}=2\pi\cdot1.05$\,MHz (see Fig. \ref{fig:Intro}d) between the cavity resonance and the first sidepeak. With larger pump powers the frequency spacing between cavity resonance and thermal sidepeak increases as the self-oscillation frequency grows. Additionally, a weak probe signal is applied for gain and noise temperature measurements. The probe frequency is swept over the whole sidepeak and the spectrum is recorded using a heterodyning spectrum analyser. We also measured the average shift of the driven cavity resonance \mw{$f_{\mathrm{r, shift}}$} using a vector network analyzer and found $\mw{f_{\mathrm{r, shift}}} - f_0= 2.3$\,MHz, which corresponds to \mw{the time-averaged temperature of graphene $\langle T_\mathrm{e}\rangle\simeq 2.4$\,K}.

The effective area of normal graphene part is $A_\mathrm{S}=1.8$ $\,\mu\mathrm{m}^2$ below the self-oscillation threshold; in the low $Q$ state, we expect the whole graphene area to dissipate energy and the effective area would become $A=40$\,$\mu\mathrm{m}^2$, but this was neglected in the analysis. Using the theoretical expression for the heat capacity we obtain $C \simeq 2\times10^{-10}(T/K)$\,J/Km$^2$ (for the chemical potential $\mu=30$\,meV), while for the electron-phonon heat flow $P_\mathrm{ep}= A_\mathrm{S} \Sigma T^3$, we can take the coupling coefficient $\Sigma=2.2\,$W/(K$^3$m$^2$) as determined from the self-oscillation period \cite{Haque2022}. Hence, the time constant $\tau=C/G_\mathrm{ep}$ for the thermal relaxation of graphene electrons via the electron-phonon coupling conductance $G_\mathrm{ep} = 3\Sigma$ is given by $\sim 25\mathrm{ps}/\,(T/K)$ in our device. This yields that, at $T=1.5 - 2$\,K, the Joule heat $V_2^2$ generation of the second harmonic temperature modulation at $2f_\mathrm{p}$ is significant in graphene for pumping frequency of 5\,GHz ($1/2\omega_\mathrm{p} \simeq 15$\,ps).  

\section{Results \& Discussions}

Fig. \ref{fig:Intro}d displays the emission spectrum as a function of pumping power $P_\mathrm{p}$ close to the threshold power $P_{\mathrm{th}}$. The emission spectrum includes more and more sidebands when $P_\mathrm{p}$ is increased. Already at $P_\mathrm{p}-P_{\mathrm{th}}=0.25$ dBm, the spectral peak sequence indicates that the time dependence $N_\mathrm{r}(t)$ is close to a sawtooth pattern. Close to the threshold, fluctuations smear the sawtooth pattern and only nearly sinusoidal modulation remains in the time dependence of $N_\mathrm{r}(t)$. Fig. \ref{fig:Mixing}b illustrates how the probe signal, applied at $f_\mathrm{p} + f_{\mathrm{T}}$ in addition to the pump, results in nearly symmetric signal and idler frequencies in our device. Both signal and idler signals are amplified, and they are seen as sharp peaks on top of broad background given by phase diffusion due to thermal self-oscillations. The basic structure of the measured emission spectra is verified by Fourier transform of the simulated time traces displayed in Fig. \ref{fig:TimeDependence}.

\begin{figure}[]
\centering
\includegraphics[width=0.99\linewidth]{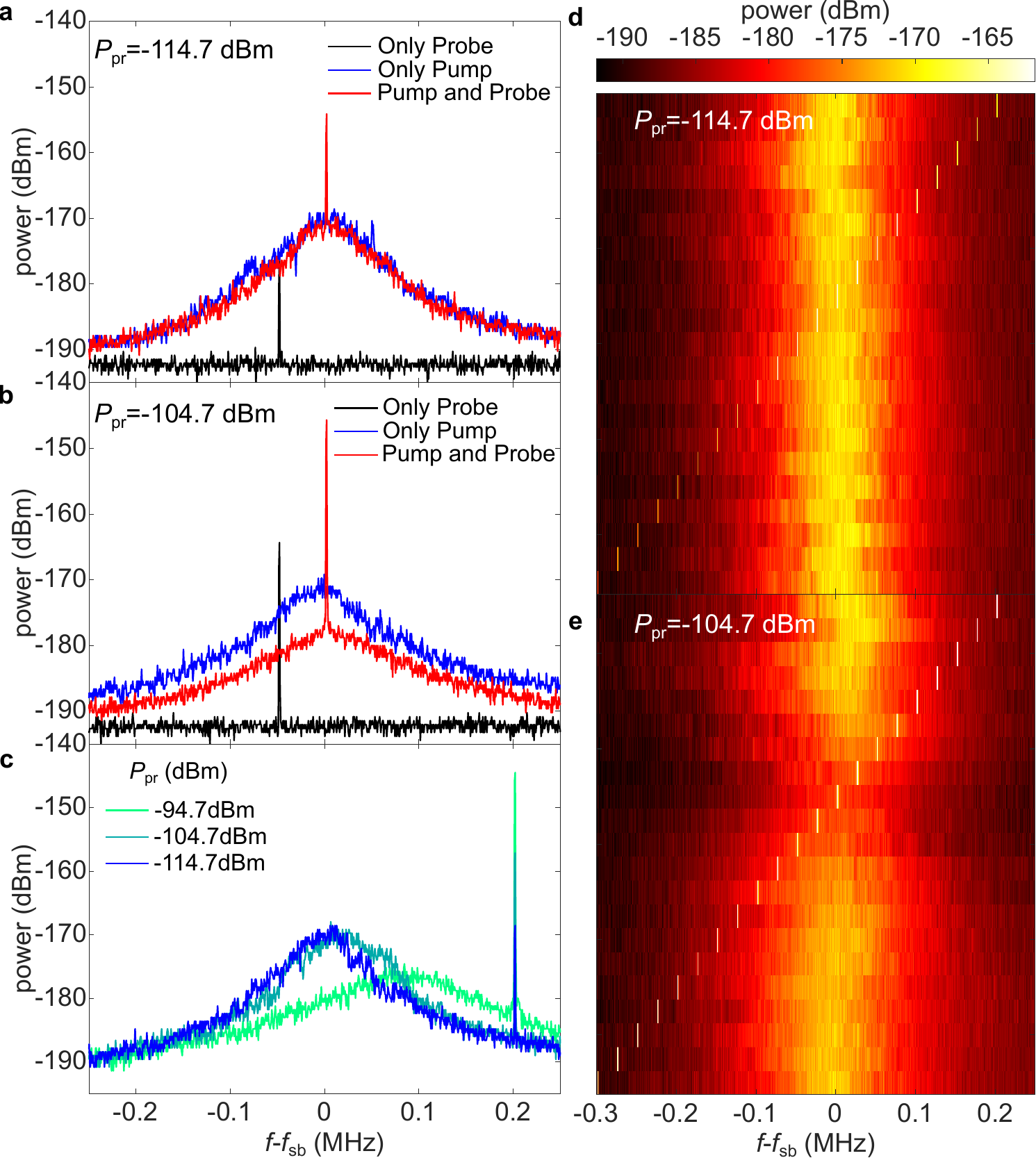}
\caption{\mw{Spectral power for the probe signal (in dBm) and for the noise background (in dBm/Hz).} \mw{a)+b) Zoom on first sidepeak for two probe powers $P_{\mathrm{pr}}$ with different pump and probe  constellations. The sideband frequency is $f_\mathrm{sb}=5.38305$\,GHz. c) Spectra with $f_{\mathrm{pr}}$ fixed near the resonance. At higher probe powers $P_{\mathrm{pr}}$ the sideband frequency shifts due to interlocking. d)+e) Emission spectra across the side peak resonance at $f_\mathrm{p}+f_\mathrm{T}$ as a function of frequency at two probe powers ($-114.7$\,dBm and $-104.7$\,dBm) measured at the pump power $-53.15$\,dBm. The probe frequency $f_{\mathrm{pr}}$ is stepped across the side peak resonance. The value of $f_{\mathrm{pr}}$ can be read from the thin bright lines on the emission spectra. At higher probe power, in frame e), injection locking of the side peak frequency to $f_{\mathrm{pr}}$ is visible when $f_{\mathrm{pr}}$ is close to $f_\mathrm{p}+f_\mathrm{T}$. The injection locking is seen as lowering in the emission power over the side peak lobe, as indicated by the  color scale for emitted power at the top.}
}
\label{fig:LockingPumpProbe}
\end{figure}

Fig. \ref{fig:LockingPumpProbe} summarizes the operation of our pumped parametric device when probe frequency is varied across the upper sideband frequency; the self-oscillation frequency stays here fixed at $f_{\mathrm{T}}=1.05$\,MHz. The right maps in Fig. \ref{fig:LockingPumpProbe} display color-coded sidepeak spectra with the different probe frequencies $f_{\mathrm{pr}}=f_\mathrm{p} + f_{\mathrm{T}}$ appearing as a thin white line in each row of the map. Results obtained at two different probe powers $P_{\mathrm{pr}}=-104.7$\,dBm and $-114.7$\,dBm are compared. Figs. \ref{fig:LockingPumpProbe}a and \ref{fig:LockingPumpProbe}b display spectra (cuts of the respective maps in Figs. \ref{fig:LockingPumpProbe}d and \ref{fig:LockingPumpProbe}e) when the probe signal is on top of the sidepeak (red lines). Additionally, traces without either pump or probe are shown (blue+black line) to highlight the difference. The stronger probe signal produces injection locking around the resonance over a frequency span of $\sim 0.125$\,MHz (see Fig. \ref{fig:LockingPumpProbe}b). This injection-locking span is the range over which the amplifier yields the largest gain and lowest noise temperature $T_\mathrm{N} \simeq 1.4$\,K. \mw{Fig. \ref{fig:LockingPumpProbe}c clearly shows the influence on the sidepeak due to injection locking for higher probe powers. \mw{Additional data measured at $P_{\mathrm{pr}}=-94.7$\,dBm is presented in \mw{Sect. II of} the SI, reaching the lowest value of $T_\mathrm{N} \simeq 1.1$\,K. Owing to the injection locking dynamics, it is hard to determine the actual saturation power of our device but it can be estimated to be in the range of $-105 \dots -100$ dBm.}}

\begin{figure}[]
\centering
\includegraphics[width=0.99\linewidth]{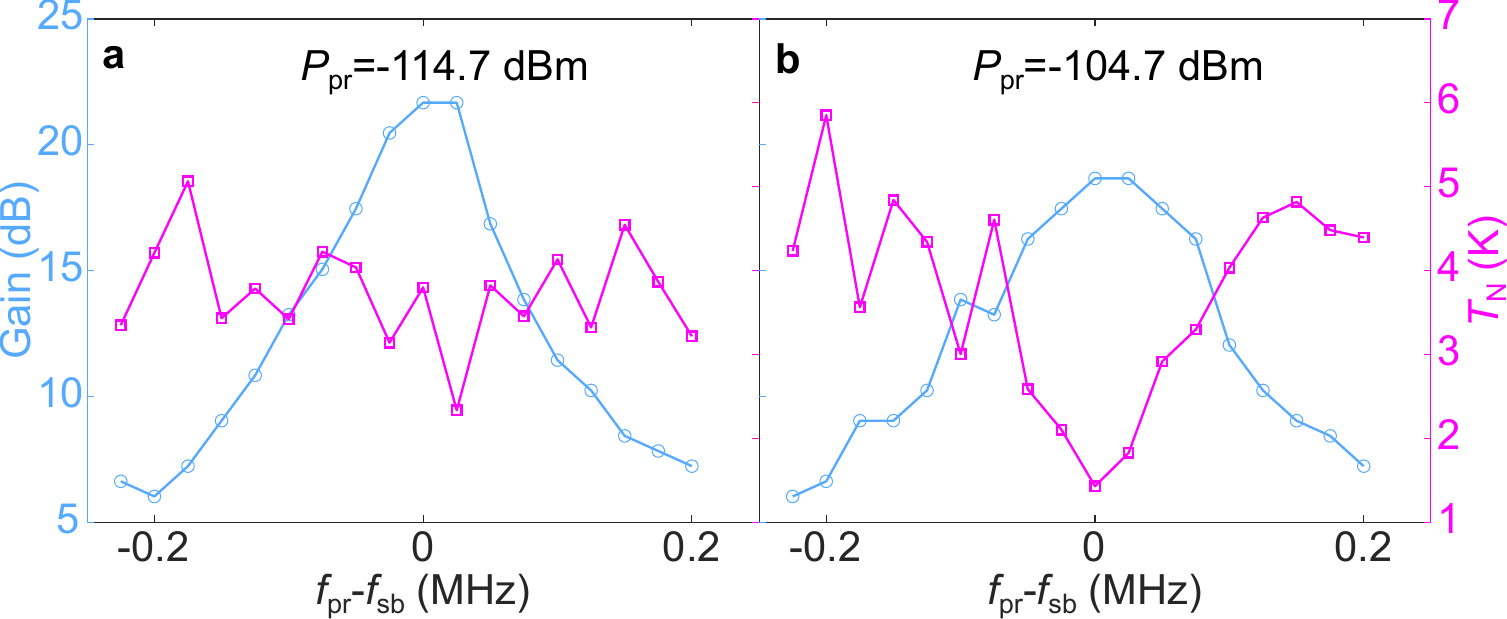}
\caption{Extracted gain (turquoise color, left scale) and noise temperature $T_\mathrm{N}$ (\mw{magenta} color, right scale) obtained from the data displayed in Fig. \ref{fig:LockingPumpProbe} for $P_{\mathrm{pr}}=-114.7$\,dBm (a) and $P_{\mathrm{pr}}=-104.7$\,dBm (b). The sideband frequency is $f_\mathrm{sb}=5.38305$\,GHz.
}
\label{fig:GainAndNoiseTemp}
\end{figure}

Fig. \ref{fig:GainAndNoiseTemp}  illustrates the gain and noise temperature performance measured for our amplifier. The operation point was selected close to the threshold power $P_{\mathrm{th}}= -53.2$\,dBm at $P_\mathrm{p}= -53.15$\,dBm. Gain and $T_\mathrm{N}$ are obtained from the measurement results displayed in Figs. \ref{fig:LockingPumpProbe}d and \ref{fig:LockingPumpProbe}e. The gain is calculated as the ratio of the maximum probe signal with and without the pump tone applied. The underlying sidepeak curves (the background level for the coherent probe signals) were fitted using Lorentzian peaks where possible; this was employed as the baseline for \mw{ extracting the signal to noise ratio. It appears, however, that the partial injection locking of the phase diffusion background varies between different measurements, which leads to substantial variation of $\pm 15$\% in the pumped signal to noise ratio and the ensuing individual noise temperature values before a stronger injection locking sets in around $P_\mathrm{pr}= -104.7$\,dBm}. The baseline for non-pumped measurement traces is obtained by a mean of points around the probe signal. The noise temperature is given by $T_\mathrm{N}=(\mathrm{SNR}_\mathrm{pump\ off}/\mathrm{SNR}_\mathrm{pump\ on})\cdot 4.3\, \mathrm{K}$, where 4.3\,K refers to the system noise temperature of the cooled HEMT preamplifier. \mw{The HEMT manufacturer specifies the $T_\mathrm{N}^\mathrm{HEMT} = 1.7$\,K around 6\,GHz measured at 5\,K bath temperature. The noise temperature of our amplifier setup is estimated by assuming approximately 1\,dB loss for each of the elements (the sample box connection, two circulators with SMA connectors, and a 60-cm-long NbTi superconducting cable) between the sample and the preamplifier (4\,dB altogether), which yields $T_\mathrm{N}^\mathrm{HEMT}=4.3$\,K. This estimate agrees well with measurements on a different sample on the same setup \mw{$T_\mathrm{N}^\mathrm{HEMT}=4.3 \pm 0.3$\,K. The error in $T_\mathrm{N}$ measured at $P_\mathrm{pr}= -104.7$\,dBm ($-114.7$\,dBm) is estimated to be $\pm 10$\% ($\pm 15$\%).}}

The measured noise temperature has a clear dip in the center of the gain curve. This dip coincides with the injection locking regime in Fig. \ref{fig:LockingPumpProbe}e. Beyond the locking regime, $T_\mathrm{N}$ is roughly the same for both probe powers. The gain for both probe powers is larger than 15\,dB over a bandwidth of 125\,kHz with a maximum gain of 18.6\,dB and 21.6\,dB, respectively. Therefore, the \mw{(voltage)} gain-bandwidth product is \mw{$\sim 1.25$\,MHz} (gain = 20\,dB).


The cavity dissipation in our amplifier  is governed, most of the time, by the dielectric/subgap losses (high $Q$ regime). The effective temperature of this dissipation is set by electronic temperature $T_e$ of the graphene resistor, i.e. the normal conductor in the GIS junctions. $T_e$ follows the pump power, and at the onset of self-oscillation, the temperature is estimated to be $\sim 2$\,K. According to our simulations, $T_e$ of graphene relaxes down to $T_e \sim 2$\,K in about 1\,ns, i.e. in $<0.2$\,\% of a typical self-oscillation cycle time. We note that this decay was calculated using relaxation only through the suspended area $A_\mathrm{S}$ (governing the subthreshold dynamics) although the dissipation in the low$-Q$ state is expected to take place over the whole area, which would result in an even faster cooling (see Sect. IV of the SI for fitting of $\omega_{\rm{T}}$ vs, $ P_\mathrm{p}$, which yields a relaxation rate of 4\,GHz). Using 20\,K for the highest temperature, time averaging yields an overall thermal contribution of $\lesssim 0.04$\,K for the effective bath noise temperature arising from the relaxation. 

Altogether, the noise temperature of our amplifier follows the weighted noise temperature $T_\mathrm{N}^w$ of the dissipation baths connected to the cavity \mw{(Our calculation is approximative, see e.g. Ref. \citenum{Parker2022} for the role of the ratio of the internal and external dissipation)}. Using $\kappa_G(T_e)$, $\kappa_{\mathrm{in}}(T_0)$,  $\kappa_{\mathrm{out}}(T_0)$, and $\kappa=\kappa_G(T_e)+\kappa_{\mathrm{in}}(T_0)+\kappa_{\mathrm{out}}(T_0)$, the effective $T_\mathrm{N}^w$ can be written as \footnote{The contribution of two level states, if any,  is basically included in the graphene dissipation term $\kappa_G(T_e)$ where they contribute to the maximum quasiparticle resistance as $T \rightarrow 0$. } $T_\mathrm{N}^w= \left(\kappa_G(T_e)T_e+\kappa_{\mathrm{in}}(T_0)T_0+\kappa_{\mathrm{out}}(T_0)T_0\right)/\kappa$. Using $\kappa_G(T_e) \simeq 4\kappa_{\mathrm{out}}$ and $T_e (T_0)=2.4$\,K (0.02\,K), we obtain $T_\mathrm{N}^w\simeq 1.9$\,K which is close to the measured noise temperature of our amplifier (see Fig. \ref{fig:GainAndNoiseTemp}). \mw{For operation near the quantum limit, a noise analysis along the lines of Ref. \citenum{Parker2022} should be done.}

In the mechanical amplifier of Ref. \onlinecite{Massel2011}, the mechanical mode acts as an idler frequency. Therefore, the number of thermal quanta in the mechanics influences directly the noise temperature of the amplifier, and cooling of the mechanical mode down to $T << \hbar \omega_m/ k_B \sim 1 $\,mK is needed to reach quantum limited operation of the amplifier. In our amplifier, however, the idler frequency is close to cavity frequency, and only the temperature of the microwave cavity is important for the noise properties of the amplifier. For quantum limited operation, we suggest that the non-linear part of the SIN capacitance is pumped using its voltage dependence instead of the temperature dependence, i.e., the device should be constructed using a small gap superconductor, such as aluminum, and the device should be operated below the self-oscillation threshold. \mw{Amplifier operation based on 3-wave or 4-wave mixing are both possible.}

There is uncertainty coming from the Poisson statistics of the cavity population. Typical cavity populations amount to $10^6 - 10^7$ quanta in our experiments. Thus, the uncertainty is approximately $ 1-3 \times 10^3$ quanta in the cavity. Assuming that the thermal switching takes place at exact classical thresholds, the Poissonian noise will only influence the time of thermal self-oscillation periods, which will acquire a relative variation of $\sim 10^{-3}$ on short time scales. This would suggest a line width of $1$\,kHz for the emission side peak at 1\,MHz self-oscillation frequency, but the observed line widths are $10-100$ times larger (e.g. $\sim 100$\,kHz in Fig. \ref{fig:Mixing}b). This indicates that phase diffusion in the parametric response plays a significant role \cite{Adler1946,Wustmann2019}. In fact, a clear reduction of the side peak phase diffusion can be achieved by increasing the probe signal at the side peak frequency (see Fig. \ref{fig:LockingPumpProbe} \mw{and  Sect. II of the SI}). Consequently, the operation of our device can be improved if it were to be used for single frequency measurements, such as null detection in bridge type of rf circuits.

In summary, we have experimentally realized a temperature driven amplifier based on a strongly nonlinear graphene SIGIS junction coupled to a microwave cavity. Amplification is achieved by driving the device into a thermal cycle with high and low dissipation regimes with a cycle time on the order of MHz. The temperature dependence leads to 2$\omega_\mathrm{p}$ pumping which modulates an effective capacitance \mw{and resistance} of the junction leading to parametric down conversion amplifying both the signal and idler frequencies located at thermal sidebands. \mw{For larger probe powers a maximum gain of $21.6$\,dB over a range of $70$\,kHz is achieved. The noise temperature at this power is $T_\mathrm{N}=2.3\,$K. At lower probe powers a noise temperature of $T_\mathrm{N}=1.4$\,K is reached, while a gain of $18.6$\,dB over 125\,kHz is obtained.} Parameter optimisation suggests that even lower noise temperatures are possible by using, for example, lower gap superconductors. \mw{Besides improved gas sensor applications \cite{Kamada2021} with integrated preamplifier, we propose}  the use of nonlinear \mw{resistance and capacitance} of a SIN junction for \mw{constructing generic parametric amplifiers, active mixers and tunable bolometers}.

\section*{Acknowledgments}

Critical remarks by Mark Dykman and clarifying correspondence with him are acknowledged. We are grateful to Manohar Kumar, Sorin Paraoanu, and Weijun Zeng  for fruitful discussions. This work was supported by the Research Council of Finland projects 341913 (EFT), and 312295 \&  352926 (CoE, Quantum Technology Finland) as well as by ERC (grant no. 670743). The research leading to these results has received funding from the European Union’s Horizon 2020 Research and Innovation Programme, under Grant Agreement no 824109 (EMP). MTH acknowledges support from the European Union’s Horizon 2020 Programme for Research and Innovation under Grant Agreement No. 722923 (Marie Curie ETN - OMT). The experimental work benefited from the Aalto University OtaNano/LTL infrastructure. This work was also supported by Ministry of Education and Culture in Finland via Aalto University's MEC Global program.

\section*{Appendix A}
\begin{table} [h!]
    \mw{\caption{Definitions of all used frequencies.}
    \vspace{10pt}
    \centering
    \begin{tabular}{ r | p{7cm} }
        symbol & description  \\ 
        \hline
        $f_\mathrm{p}$ & Pump frequency of the cavity, set to $f_\mathrm{r}$(20\,mK).  \\  
        $f_\mathrm{pr}$ & Probe frequency: pump and probe are used in the input signal to the cavity $v_{\mathrm{in}}(t) = v_p \sin(\omega_\mathrm{p} t) + v_{\mathrm{pr}}\sin(\omega_\mathrm{pr}t)$.  \\
        $f_\mathrm{T}$ & Frequency of the thermal self-oscillation of the cavity-SIN system.  \\  
        $f_{0}$ & Un-pumped cavity frequency at 10\,mK,  $f_{0}=f_\mathrm{r}(10\,\mathrm{mK})$.  
        \\
        $f_\mathrm{r}$ & Temperature dependent resonance frequency of the cavity $f_\mathrm{r}(T)$.  \\
        $f_\mathrm{r,shift}$ & Temperature dependent resonance frequency of the cavity $f_\mathrm{r}(T)$ shifted by pumping power $P_{\mathrm{p}}$.  \\
        $f_\mathrm{sb}$ & Centre frequency of the sideband arising due to thermal oscillation at $f_\mathrm{p}+f_\mathrm{T}$. \\ 
        $\delta f_\mathrm{eff}$ & Frequency shift induced by the effective capacitance modulation $\delta C_\mathrm{eff}$.
    \end{tabular}
    \label{tbl:frequencies}}
\end{table}

\onecolumngrid


\twocolumngrid
\bibliography{references}

\end{document}